\documentclass[10pt]{iopart}
\usepackage[utf8]{inputenc}
\usepackage[english]{babel}

\usepackage{graphicx}
\usepackage{dcolumn}
\usepackage{amssymb}
\usepackage{lmodern}
\usepackage{longtable}
\usepackage{units}
\usepackage{bm}

\newcommand{\TC}{T_{\rm C}} 
\newcommand{\Te}{T_{\rm el}} 
\newcommand{\Tl}{T_{\rm ph}} 
\newcommand{\Gel}{G_{\rm el-ph}} 
\newcommand{\Ce}{C_{\rm el}} 
\newcommand{\GammaCe}{\gamma_{\rm \Ce}} 
\newcommand{\Cl}{C_{\rm ph}} 
\newcommand{\Tauth}{\tau_{\rm th}} 
\newcommand{\TauP}{2 \sigma^2} 
\newcommand{\kB}{k_{\rm B}}
\newcommand{\damping}{\alpha}
\newcommand{\Jij}{J_{ij}} 
\newcommand{\aO}{a_0} 
\newcommand{\TO}{T_{0}} 
\newcommand{\PO}{P_0} 
\newcommand{\tO}{t_0} 
\newcommand{\di}{\rm{d}}
\newcommand{\argu}[1]{\left(#1\right)}
\newcommand{\set}[1]{\left\lbrace#1\right\rbrace}
\newcommand{\abs}[1]{\left\vert#1\right\vert}
\newcommand{\avg}[1]{\left\langle#1\right\rangle}

\begin{document}

\title{Role of temperature-dependent spin model parameters in ultra-fast magnetization dynamics}
\author{A.\ De\'ak$^{1,2}$, D.\ Hinzke$^3$, L.\ Szunyogh$^{1,2}$, U.\ Nowak$^3$} 
\address{$^1$Department of Theoretical Physics, Budapest University of Technology and Economics, Budafoki \'ut 8., HU-1111 Budapest, Hungary}
\address{$^2$MTA-BME Condensed Matter Research Group, Budapest University of Technology and Economics, Budafoki \'ut 8., HU-1111 Budapest, Hungary}
\address{$^3$Fachbereich Physik, Universit\"{a}t Konstanz, D-78457 Konstanz, Germany}

\ead{adeak@phy.bme.hu}

\date{\today}

\begin{abstract}
In the spirit of multi-scale modelling magnetization dynamics at elevated temperature is often simulated  in terms of a spin model where the model parameters are derived from first principles. While these parameters are mostly assumed temperature-independent and thermal properties arise from spin fluctuations only, other scenarios are also possible. Choosing bcc Fe as an example, we investigate the influence of different kinds of model assumptions on ultra-fast spin dynamics, where following a femtosecond laser pulse a sample is demagnetized due to a sudden rise of the electron temperature. While different model assumptions do not affect the simulational results qualitatively, their details do depend on the nature of the modelling.
\end{abstract}



\ioptwocol %

\section{Introduction} 
The temperature dependence of macroscopic physical quantities is mostly due to thermal fluctuations of the microscopic degrees of freedom, which---for magnetic systems---is the atomic spin magnetic moment, in the following for simplicity called spin. 
The Hamiltonian of the system is normally parameterized in form of a spin model where the model parameters are assumed temperature-independent while the spins---in the classical limit---are allowed to fluctuate with their magnitude kept constant. For a given material the model parameters can be derived from first principles, mostly relying on the famous approach of Liechtenstein \emph{et al}.\cite{LiechtensteinJMMM87} Different related methods have been developed  in the past suitable to treat correlated systems \cite{KatsnelsonPRB2000,KatsnelsonEPJB2002}, relativistic effects \cite{udvardiPRB03,ebertPRB09} or both of them \cite{KatsnelsonPRB2010,SecchiAnnPhys2015}.  Quite recently, an approach to calculate magnetic interactions under non-equilibrium conditions has also been developed \cite{SecchiAnnPhys2013}.

Usually, first-principles electronic-structure calculations and the determination of exchange parameters are performed for zero temperature. In a second step, these exchange parameters are used for a spin model to calculate thermal properties via Monte Carlo simulations or Langevin dynamic simulations based on the stochastic Landau--Lifshitz--Gilbert equation of motion \cite{nowakBOOK07}, i.e.\ the same exchange parameters are used to describe magnetic systems at elevated temperatures. However, the question arises how far the parameters of the spin model itself are temperature-dependent due to thermally-induced changes of the electronic structure or the temperature-dependent Weiss field surrounding the atoms. 
Hence, new approaches to obtain temperature-dependent spin model parameters computed from first principles have been proposed 
either in terms of Fermi-Dirac statistics  \cite{chimataPRL12} or considering non-collinear spin-configurations due to thermal fluctuations \cite{szilvaPRL13}.
Recently, B\"ottcher \emph{et al.} \cite{bottcherJMMM12} proposed a method to calculate the temperature dependence of Heisenberg exchange coupling constants. The magnetization (order parameter) of the system was controlled using an Ising type approach, by mixing up- and down- magnetic moments in different concentrations $c$.  The temperature was then adjusted to $c$ by fitting the magnetization from Monte Carlo simulations using the calculated exchange coupling constants to that obtained from \emph{ab initio} calculations. 

To further develop this concept,  we present an \emph{ab initio} model of the temperature dependence of the Heisenberg exchange coupling constants $J_{ij}$ as well as the atomic magnetic moments $\mu$ based on the  scheme of disordered local moments (DLM) \cite{gyorffyJPFMF85}. The relativistic extension of this theory \cite{stauntonPRL04,stauntonPRB06} allows for a direct link of the magnetization (temperature) to the electronic structure, thus to the calculated exchange constants. Moreover, as in Reference~\cite{chimataPRL12}, we use the Fermi--Dirac distribution to include effects of thermal electronic excitations.
As a model system we consider bcc Fe. Starting with equilibrium properties we expand our investigation to the case of ultra-fast magnetization dynamics where---following a strong excitation with a femtosecond laser pulse---the sample first demagnetizes on a sub-picosecond time scale and then recovers its magnetization on a larger time scale. This effect was discovered by Beaurepaire \cite{beaurepairePRL96}, a work which has inspired the field of ultra-fast spin dynamics with potential applications in data storage. Since the laser pulse leads to a strong increase of the electron temperature, an influence of temperature-dependent variations of the spin model parameters can be expected and will be studied in the following. 

Our work is structured as follows: in the next chapter we will introduce our methods, first-principles calculations to derive the model parameters as well as Langevin dynamics for the thermodynamic calculations. Then we discuss the temperature dependence of the model parameters that we derived from first principles. In the following two sections we discuss the results from our thermodynamic calculations, first the equilibrium properties and then the spin dynamic behavior as triggered by a short laser pulse.

\section{Theory: From First Principles to Atomistic Spin Model} 

First, we compute the
electronic structure of the magnetic system at finite temperatures in terms of the relativistic disordered local moment (RDLM) scheme \cite{gyorffyJPFMF85,stauntonPRL04,stauntonPRB06} implemented in the screened Korringa--Kohn--Rostoker method.\cite{KKRBOOK05} The RDLM scheme relies on the adiabatic approximation where the slow spin degrees of freedom are decoupled from the fast (electronic) degrees of freedom and the configuration of the local moments can be described by a set of unit vectors $\left\lbrace \mathbf e\right\rbrace=\left\lbrace \mathbf e_1, \mathbf e_2, \dots \mathbf e_N\right\rbrace$. The RDLM theory describes the fluctuations of the finite-temperature system in terms of single-site probabilities,
\begin{equation}
\mathcal{P}\argu{\set{\mathbf e}}=\prod_i  P_{i}\argu{\mathbf e_{i}},
\end{equation}
inherently providing a local mean-field description of spin disorder. In the framework of the single-site coherent potential approximation (CPA), a magnetically ordered coherent medium described by the $t$-matrices $\underline t_{c,i}$ are sought for at every given energy $\varepsilon$ (not noted explicitly), and the self-consistency condition for the corresponding scattering path operator (SPO), \cite{KKRBOOK05}
\begin{equation}
\underline{\underline \tau}_c=\left(\underline{\underline t}_c^{-1}-\underline{\underline G}_0\right)^{-1}, \label{eq:tauc}
\end{equation}
reads as
\begin{equation}
\underline\tau_{c,ii}=\int \bigl\langle \underline\tau_{ii} \argu{\set{\mathbf e}}\bigr\rangle_{\mathbf e_{i}} P_{i}\argu{\mathbf e_{i}}\,\mathrm d^2 e_{i},
\end{equation}
where $\bigl\langle \underline\tau_{ii} \argu{\set{\mathbf e}}\bigr\rangle_{\mathbf e_{i}}$ denotes the partial average of the scattering path operator with the spin direction fixed at site $i$. Note that quantities with one and two underlines denote matrices with angular momentum indices and with combined site -- angular momentum indices, respectively, while $\underline{\underline G}_0$ in \Eref{eq:tauc} stands for the free-space structure constants.  An equivalent form of the CPA condition can be given in terms of the excess scattering matrices $\underline X_{i}$,
\begin{equation}
\underline X_{i}\!\left(\mathbf e_{i}\right)=\left[\left(\underline t_{c,i}^{-1}-\underline t_{i}^{-1}\!\left(\mathbf e_{i}\right)\right)^{-1} - \underline \tau_{c,ii}\right]^{-1},
\end{equation}
as
\begin{equation}
\avg{\underline X_{i}\argu{\mathbf e_{i}}}=\int P_{i}\argu{\mathbf e_{i}} \underline X_{i}\!\left(\mathbf e_{i}\right)\, \mathrm d^2 e_{i} =\underline 0 .
\end{equation}

The single-site probabilities $P_{i}\argu{\mathbf e_{i}}$ can be determined self-consistently,\cite{deakPRB14} and the finite-temperature orientational distribution can be characterized using the dimensionless average magnetization
\begin{equation}
m_{i}=\abs{\avg{\mathbf e_{i}}}=\abs{\int \mathbf e_{i} \,P_{i}\argu{\mathbf e_{i}} \mathrm d^2e_{i}}.
\end{equation}
This quantity also plays the role of an order parameter in a homogeneous system, with $m=1$ corresponding to ferromagnetic order and $m=0$ to the paramagnetic state.

Besides considering temperature-induced transversal spin fluctuations, we also include the effect of finite electronic temperature, $T_{\rm el}$, by using the Fermi-Dirac distribution to calculate averages over electronic states. Such an extension of the Density Functional Theory was founded by Mermin, \cite{Mermin-1965} see also Reference~\cite{KueblerBook}. Applying this approach to ferromagnets  obviously leads to too high Curie temperatures \cite{Gunnarsson-1976, KueblerBook} due to neglected orientational fluctuations of the local moments. 
In our implementation of the self-consistent RDLM method we  consider both mechanisms for temperature dependence and calculate the electronic and magnetic structure of the system as a function of magnetic disorder and electronic temperature simultaneously. 

Starting from a self-consistent-field calculation performed for a given order parameter and electronic temperature we can obtain spin model parameters by combining the relativistic torque method with the RDLM reference state. The relativistic torque method\cite{udvardiPRB03} extracts exchange interactions from the electronic structure by considering infinitesimal rotations of pairs of spins with respect to the ordered ground state. We can generalize this approach by considering the perturbation of two spins (say, at sites $i$ and $j$) immersed in the fluctuating system, and assessing the corresponding change in the two-site restricted thermodynamical average of the grand potential, $\avg{\Omega\argu{\set{\mathbf e}}}_{\mathbf e_i \mathbf e_j}$. The part of this quantity which depends simultaneously on $\mathbf e_i$ and $\mathbf e_j$, is given by\cite{szunyoghPRB11} 
\begin{eqnarray}
\fl \avg{\Omega\argu{\set{\mathbf e}}}_{\mathbf e_i \mathbf e_j}\approx  &-&\frac{1}{\pi} \mathrm{Im} \int\limits_{-\infty}^\infty f\argu{\varepsilon;\mu} \times\\
&\displaystyle\sum\limits_{k=1}^\infty& \frac{1}{k} \Tr\left[\underline X_i\argu{\mathbf e_i} \underline \tau_{c,ij} \underline X_j\argu{\mathbf e_j}\underline\tau_{c,ji}\right]^k \mathrm d\varepsilon, \nonumber
\end{eqnarray}
where $f\argu{\varepsilon;\mu}$ is the Fermi function with the chemical potential $\mu$. 
Note that the so-called backscattering terms\cite{Butler-1985} explicitly containing sites other than $i$ and $j$  are neglected in the above expression.
Expanding the two-site averaged grand potential up to second order in the change in inverse $t$-matrices ultimately leads to the expression for the two-site derivative:
\begin{eqnarray}
\fl \frac{\partial^2 \avg{\Omega\argu{\set{\mathbf e}}}_{\mathbf e_i \mathbf e_j}}{\partial \phi_{1i} \partial \phi_{2j}}=-\frac{1}{\pi}\mathrm{Im}\Tr \int\limits_{-\infty}^\infty f\argu{\varepsilon;\mu} \times\qquad  \label{eq:rtm-rdlm}\\
\qquad \frac{\partial \underline t_i^{-1}\argu{\mathbf e_i}}{\partial \phi_{1i}} \avg{\underline \tau_{ij}}_{\mathbf e_i,\mathbf e_j} \frac{\partial \underline t_j^{-1}\argu{\mathbf e_j}}{\partial \phi_{2j}} \avg{\underline \tau_{ji}}_{\mathbf e_i,\mathbf e_j} \mathrm d\varepsilon.\nonumber
\end{eqnarray}
In the spirit of the relativistic torque method, the derivatives are taken with respect to infinitesimal rotational angles, $\phi_{1i}$ and $\phi_{2j}$, around two orthogonal unit vectors perpendicular to the ground-state orientation of the spins at sites $i$ and $j$. Combining these
derivatives for three orthogonal directions of the overall average magnetization of the ferromagnetic system, the exchange coupling tensor $J_{ij}^{\alpha \beta}$ $(\alpha, \beta=x,y,z)$ can be derived. \cite{udvardiPRB03} While in this way it is possible to obtain relativistic contributions to the exchange interactions, namely, two-site anisotropies and Dzyaloshinskii--Moriya interactions, in this work we only consider isotropic Heisenberg interactions identified as $J_{ij}=(J_{ij}^{xx}+J_{ij}^{yy}+J_{ij}^{zz})/3$.

We note that the non-relativistic form of \Eref{eq:rtm-rdlm} was used by B\"ottcher \emph{et al.}\cite{bottcherJMMM12} to compute magnetization-dependent isotropic exchange interactions. In their approach, atoms with up- and down-moments were distributed randomly  in the framework of the so-called partial DLM approach. Instead of performing a proper thermodynamic average, the concentration of the components was fitted to the temperature-dependent average magnetization obtained from Monte Carlo simulations.  We emphasize that by employing self-consistent orientational distributions for the local moments and by incorporating finite electronic temperature now we can elaborate on the subject of finite-temperature effects on the exchange interactions.

In our hierarchical multiscale approach, the computed parameters, namely, the magnitude of the local magnetic moment $\mu$ and the exchange constants $\Jij$ are used for numerical simulations based on an atomistic Heisenberg spin Hamiltonian.  We consider thereto classical spins $\mathbf{S}_i = \boldsymbol{\mu}_i/\mu$, $\boldsymbol{\mu}_i$ being the local magnetic moment at site $i$,
with the following Heisenberg spin Hamiltonian
\begin{eqnarray}
{\cal H} =& -&\sum\limits_{i \ne j} \Big( \frac{\Jij}{2} \mathbf{S}_i  \cdot \mathbf{S}_j \label{eq:generic-Hamiltonian} \\ \nonumber   &-& \frac{\mu_0 \mu^{2}}{8 \pi}  \frac{3 (\mathbf{S}_i \cdot {\mathbf n}_{ij})({\mathbf n}_{ij} \cdot \mathbf{S}_j) - \mathbf{S}_i \cdot \mathbf{S}_j} {r_{ij}^3} \Big). 
\end{eqnarray}
The first sum represents the exchange energy of magnetic moments and the second sum describes the magnetic dipole-dipole energy, in which $\mu_0$ is the vacuum permeability, ${\mathbf n}_{ij}$
and $r_{ij}$ denote the direction (unit vector) and the distance between site $i$ and $j$, respectively.

To calculate thermal properties we use Langevin dynamics, i.e.\ numerical solutions of the stochastic LLG equation of motion,  
\begin{eqnarray}  
           \frac{ (1+\damping^2)}{\gamma} \mu {\dot{\bf{S}}}_i =
              -   \mathbf{S}_i  \times  \left[ \mathbf{H}_i 
              + \damping \;   \left( \mathbf{S}_i  \times \mathbf{H}_i  \right) \right],
\end{eqnarray}
with the gyromagnetic ratio $\gamma$, and a dimensionless Gilbert damping constant $\damping$  that describes the coupling to the heat bath. Thermal fluctuations are included as an additional noise term $\zeta$ in the internal fields $\mathbf{H}_i= - \frac{\partial \mathcal{H}}{\partial  \mathbf{S}_i} + \mbox{$\bm \zeta$}_i(t)$ with  
\begin{equation}
\langle \mbox{$\bm \zeta$}_i(t) \rangle = 0, \quad 
 \langle\zeta_{i\eta}(0) \zeta_{j\theta}(t) \rangle =\frac{2 \kB \Te \damping \mu}{\gamma} \delta_{ij}\delta_{\eta\theta} \delta(t), 
\end{equation}
where $\eta,\theta$  are Cartesian components.  All algorithms we use are described in detail in Reference~\cite{nowakBOOK07}. 

As described above the exchange interactions are computed as a function of order parameter $m$ and electronic temperature $T_{\rm el}$. In the simulations we therefore have to evaluate the dimensionless average magnetization $m=\abs{\avg{\mathbf S_i}}$ and identify it with the order parameter used in the RDLM calculations, which together with the electronic temperature specify the exchange couplings to be used in the following time steps. Similarly, the magnitude of the local moment is suitably updated during a simulation according to the order parameter and the electronic temperature.

Note that the exchange interactions are incorporated in our atomistic spin dynamics simulations via the Fast Fourier transformation method (see Reference~\cite{hinzkeJMMM00} for more details).  As a side effect, we are able to calculate the dipolar interaction without any additional computational effort so we take them into account although they will not  influence our results much.

\section{Temperature dependence of magnetic moments and exchange interactions} 

\begin{figure}[]
\hspace*{-0.5cm}
\begin{centering}
\includegraphics[scale=0.7, angle = 0]{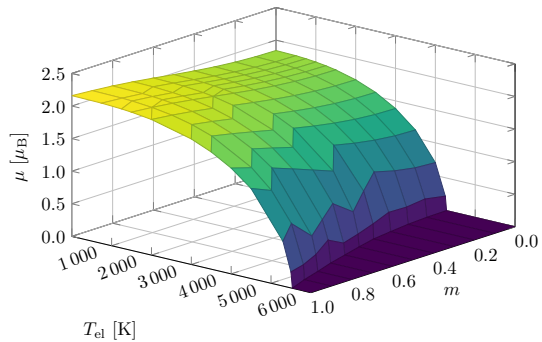}
\end{centering}
\caption{Local magnetic moment $\mu$ as function of order parameter $m$ and electron temperature $\Te$ as calculated from first principles.}
\label{fig:magneticmoment}
\end{figure}

First we performed self-consistent RDLM calculations of bulk Fe by varying the degree of spin disorder and the electronic temperature independently. For the lattice constant of the bcc lattice we used $a=5.27 \; \aO$ corresponding to the LSDA total energy minimum obtainable with ASA-KKR.\cite{mjw-book-1978} 
The dependence of the local magnetic moment $\mu$ of bulk bcc Fe as a function of  $m$ and  $\Te$ is shown in Fig.\ \ref{fig:magneticmoment}.\footnote{Surface plots in this paper use the \texttt{viridis} colormap of the \texttt{matplotlib} library, see Reference~\cite{Hunter-2007}.} At zero electronic temperature there is a reduction of the local Fe moment of almost $15\%$ in the paramagnetic state compared to the ground state, suggesting that the local moment of bcc Fe is not as rigid as it is often assumed.\cite{deakPRB14} It is thus indeed preferable to probe its behavior beyond the usual Heisenberg model, by incorporating the effect of longitudinal spin-fluctuations into the spin model. 

In good agreement with calculations of Chimata \emph{et al.}\cite{chimataPRL12} the local moment is quite stable for about $\Te \le 2000$~K and for higher electron temperatures it rapidly drops. The Stoner--Curie temperature $T_\mathrm{SC}$ at which the local moment of Fe vanishes is about 5500~K in case of $m=1$.  This value is clearly lower than that found by Chimata \emph{et al.}, $T_\mathrm{SC}=6030$~K. The most probable reason for this deviation is that Chimata \emph{et al.}\ used the experimental lattice constant $a=5.41 \; \aO$ and for higher atomic volumes the local moment of Fe is more stable against thermal excitations than for lower atomic volumes.

\begin{figure}
\hspace*{-0.5cm}
\begin{centering}
\includegraphics[scale=0.7, angle = 0]{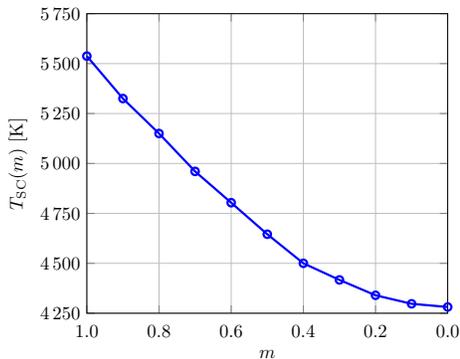}
\end{centering}
\caption{Calculated dependence of the Stoner--Curie temperature  $T_\mathrm{SC}$ on order parameter $m$.}
\label{fig:stonercurie}
\end{figure}

A new feature that can be inferred from Fig.~\ref{fig:magneticmoment} is that the  Stoner--Curie temperature decreases with decreasing order parameter. This is clearly highlighted in Fig.~\ref{fig:stonercurie} showing a monotonic decrease of $T_\mathrm{SC}$ against increasing spin disorder.  This tendency can be anticipated from the decrease of the local moment with increasing $m$ as pointed out above.    

\begin{figure*}
\hspace*{-0.5cm}
\begin{centering}
\includegraphics[width=0.8\linewidth, angle = 0]{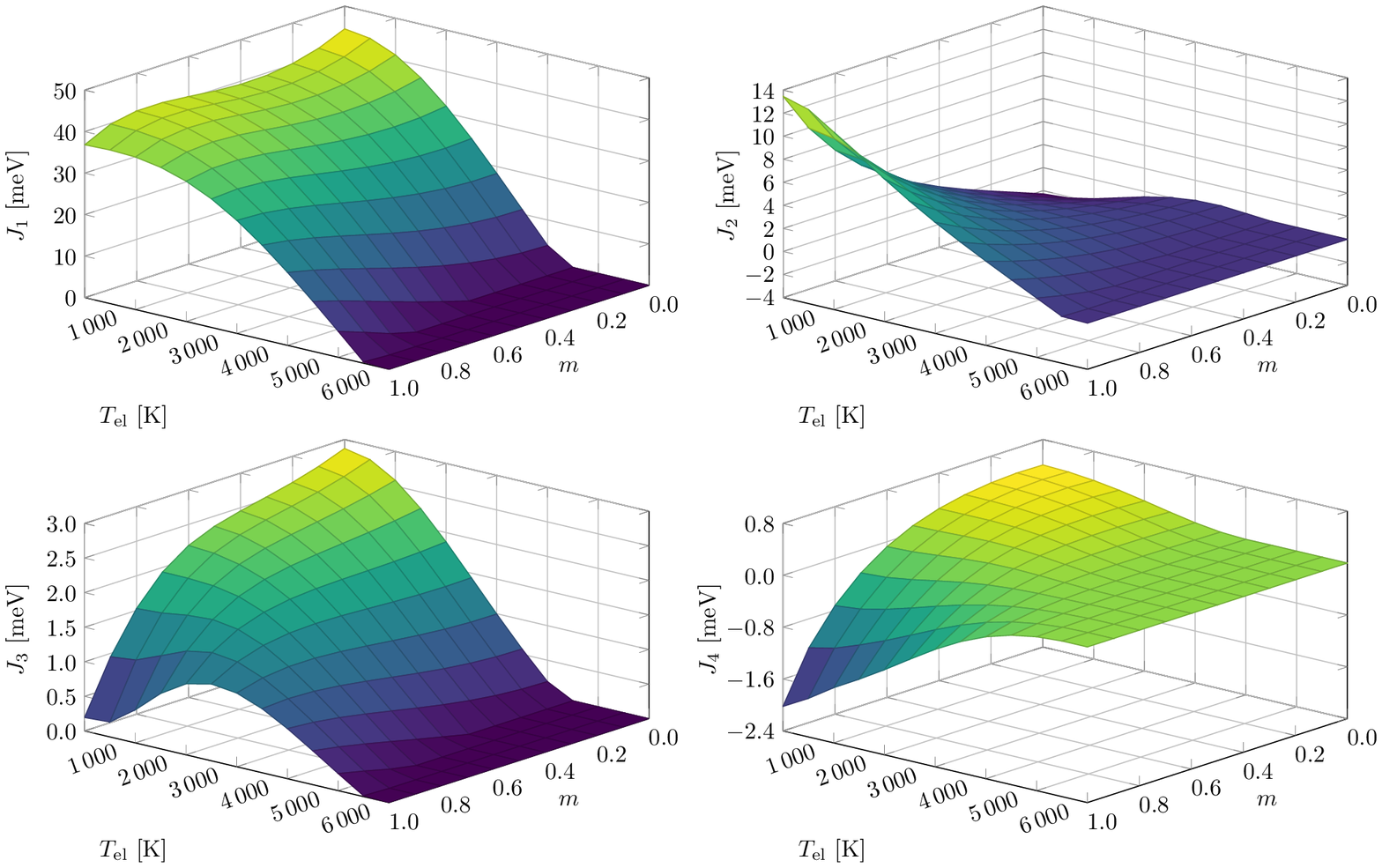}
\end{centering}
\caption{Calculated isotropic exchange constants for the first four nearest neighbors, $J_i $ ($ i=1, \dots , 4$), as function of the order parameter $m$ and electron temperature $\Te$.}
\label{fig:exchange}
\end{figure*}

Following the self-consistent calculations we also calculated exchange interactions  as described in the previous section. The isotropic interactions,  $\Jij\argu{m,\Te}$, are shown in Fig.~\ref{fig:exchange} for the first four nearest neighbors (NN). Even though the dominant first neighbor couplings are maximal for zero electron temperature, some further neighbors show interesting non-monotonic behavior as a function of $\Te$, ultimately vanishing at the respective Stoner--Curie points. At $\Te=0$~K, the dependence of the dominant exchange interaction is also non-monotonic as a function of $m$, furthermore some couplings show an enhancement towards the paramagnetic (PM) ($m=0$) limit. A similar but much larger enhancement of the first NN Fe-Fe coupling in the PM phase was found by B\"ottcher \emph{et al.}\cite{bottcherJMMM12}: taking into account a factor of one-half due to the different definition of the spin Hamiltonian, the first NN coupling in our calculation ($\sim$ 45 meV) and that of B\"ottcher \emph{et al.}\ ($\sim$ 52 meV) agree well in the disordered PM state, while in the ferromagnetic (FM) state ($m=1$) the latter one is considerably smaller ($\sim$ 22 meV) than ours ($\sim$ 38 meV).

\section{Equilibrium magnetization} 

In the LLG simulations we consider a system of size 35.86 nm $\times$ 35.86 nm $\times$ 35.86 nm with a lattice constant of 5.27 $\aO$ in the high damping limit ($\damping = 1$). The exchange couplings as described in the previous section were taken into account up to a distance of six lattice constants. We compare simulations where we either assume constant ground state spin model parameters (exchange interaction and magnetic moment) fixed at $m=1$ and $\Te=0$~K,  $m$-dependent spin model parameters at $\Te=0$~K, $\Te$-dependent parameters for $m=1$, and the parameters with full temperature ($m$- and $\Te$-) dependence.

\begin{figure}
\begin{centering}
\includegraphics[scale=0.85, angle = 0]{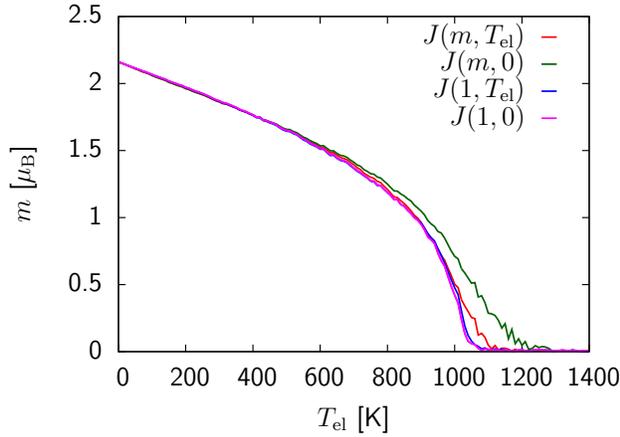}
\end{centering}
\caption{Thermally averaged magnetic moment per atom vs.\ electron temperature from Langevin dynamics simulations using spin model parameters with different types of temperature dependence (see text).}
\label{fig:equimag}
\end{figure}

Figure~\ref{fig:equimag} shows the zero-field thermally averaged magnetic moment per atom vs.\ electron temperature $\Te$ for bcc Fe comparing temperature-independent ground-state spin model parameters (for $m=1$)  with temperature-dependent spin model parameters for the respective values of the magnetization $m$. The magnetization curve is clearly affected by the temperature dependence of the exchange. The larger Curie temperatures when taking into account the $m$-dependence of the exchange coupling is due to the fact that the most important 1st NN coupling increases with decreasing oder parameter (increasing temperature), see Fig.~\ref{fig:exchange}. However, taking into account the effect of the electron temperature lowers the strength of the 1st NN coupling. For comparison, the experimental value of the Curie temperature of bcc Fe is $\TC = 1045$ K. Note however, that this experimental value should not be taken as benchmark for our calculation, since the results of the first principles calculations also depend on the assumptions made regarding the value of the atomic distance \cite{Buruzs-2008}. We have chosen the lattice constant of 5.27 $\aO$ since   it is in the vicinity of the value optimized by first principles calculations \cite{mjw-book-1978, huhnePRB98} and, fortunately, the simulated Curie temperature is close to the experimental value, see Fig.~\ref{fig:equimag}.

\section{De- and remagnetization due to a laser pulse} 

In the following, we study the dynamic reaction of the magnetization following the excitation via a fs laser pulse. This is especially interesting in the context of our first-principles calculations, since following a laser pulse the electron temperature can reach large values far above the Curie temperature, while on this short time scale the order parameter remains finite. Hence, under these strong non-equilibrium conditions, the dependence of the spin model parameters on $m$ and $\Te$ can be tested separately. 

Similar to References \cite{kazantsevaEPL08} and \cite{hinzkePRB15} we couple a spin model to a heat bath the temperature dynamics of which are calculated from a well-established two-temperature model derived by Kaganov \emph{et al}.\cite{kaganovJETP57}\ to describe the temperature evolution in our Fe sample after an excitation with a fs laser pulse with a Gaussian laser profile $P(t) = \PO \exp(-(t-\tO)^2/\TauP)$ with $\TauP = \tau^2/(4\ln(2))$. To calculate the response of the laser pulse, we use the following coupled differential equations for our two-temperature model,

\begin{eqnarray}
\Cl \frac{\di \Tl}{\di t} &=& -\Gel(\Tl\!-\!\Te) \!-\! \Cl \frac{\Tl\!-\!\TO}{\Tauth},\nonumber\\
\Ce(\Te)\frac{\di \Te}{\di t} &=& -\Gel(\Te\!-\!\Tl)\!+\!P(t),
\label{Eq-2TM}
\end{eqnarray}
where $\Te$ and $\Tl$ are the temperatures of the electronic and lattice reservoirs, $\Ce = \GammaCe \Te$ (with $\GammaCe = 670$ J/(m$^3$K$^2$)) and $\Cl = 2.2 \times 10^6$ J/(m$^3$K))  are the electronic and lattice specific heats, respectively, and $\Gel = 4.05 \times 10^{18}$  J/(sm$^3$K)) is the electron-phonon coupling constant. The time constant $\Tauth = 50$ ps describes the relaxation back to the initial temperature $\TO$. The parameters used in the model were taken from Reference~\cite{chimataPRL12}. 

\begin{figure}[h]

\begin{centering}
\includegraphics[clip, width=0.9\columnwidth]{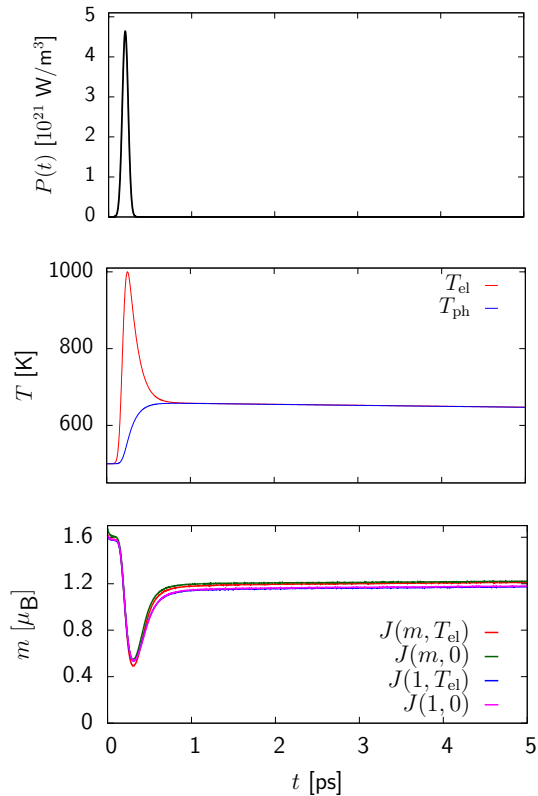}
\end{centering}
\caption{De- and remagnetization processes due to a laser pulse considering different types of spin model parameters, temperature-independent as well as temperature-dependent (bottom panel). The time-dependent Gaussian laser pulse $P(t)$ with $\PO = 4.63 \cdot 10^{21}$ W/m$^2$, the electron $T_{\rm {el}}$ as well as phonon temperature $T_{\rm {ph}}$ are shown in the upper and middle panels, respectively.}
\label{fig:Mvst_1500K}
\end{figure}

A spin system of size 17.87 nm $\times$ 17.87 nm $\times$ 17.87 nm  is simulated with a damping constant of $\damping = 0.1$. As an example, in Fig.\ \ref{fig:Mvst_1500K} the laser profile $P(t)$ as well as the electron  $\Te$ and lattice temperature $\Tl$ calculated from Eqs.\ \ref{Eq-2TM} for a laser pulse with $\PO =  4.63 \cdot 10^{21}$ W/m$^2$ and $\tau = 84$ fs are shown. 

\begin{figure}[ht]
\begin{centering}
\includegraphics[width = 0.9\columnwidth]{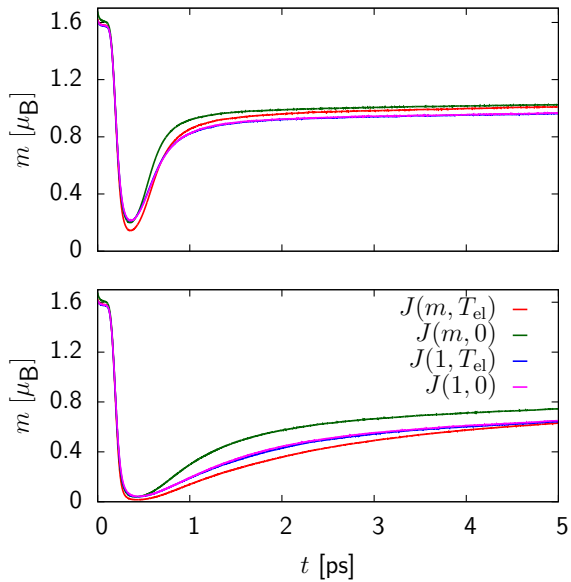}
\end{centering}
\caption{Time-dependent magnetizations as in Fig.\ \ref{fig:Mvst_1500K} for two further laser pulse powers, $7.74 \cdot 10^{21}$ W/m$^2$ (top) and $1.23 \cdot 10^{22}$ W/m$^2$ (bottom).}
\label{fig:MvsT_versP}
\end{figure}

In Figs.\ \ref{fig:Mvst_1500K} and \ref{fig:MvsT_versP} the quenching and the relaxation of the magnetization following a thermal excitation with laser pulses of three different intensities are compared. As in the equilibrium case of the previous section we consider different types of spin model parameters, assuming either temperature-independent ground-state parameters (for $m = 1$) or temperature-dependent spin model parameters for the respective time-dependent values of the order parameter $m$, with and without taking into account the time-dependent electron temperature $\Te$. 

In Fig.\ \ref{fig:Mvst_1500K} the influence of the different types of model parameters on the magnetization dynamics is rather small. This is mainly due to the fact that the laser power is assumed rather small, so that the order parameter does not vanish completely but is rather quenched by about 50\% only. Here, the effect of the electron temperature variation is also not big enough to lead to major variations of the magnetization dynamics. Furthermore, the temperature dependence of the atomic magnetic moment $\mu$ as shown in Fig.\ \ref{fig:magneticmoment} is still not visible since even the rather high electron temperatures considered here are still far below the Stoner--Curie temperature.  

This is different in Fig.\ \ref{fig:MvsT_versP}. For intermediate laser powers (upper graph) the quenching of the magnetization is slightly affected for the case where both the electron temperature-dependence of the spin model parameters, as well as their order parameter-dependence is considered (red line). Since both effects lead to a reduction of the 1st and 2nd NN exchange constants in that temperature range (see Fig.~\ref{fig:exchange}) the demagnetization is stronger. For even higher laser power the sample is completely demagnetized after the laser pulse for all model assumptions. But now the relaxation phase is affected and it is the model with $m$-dependent but $\Te$-independent exchange parameters which shows the quickest relaxation. This is due to the fact that this model has the largest 1st NN exchange parameters in that temperature range, which also leads to the highest Curie temperature in our equilibrium calculations (see Fig.~\ref{fig:equimag}).

\section{Summary}   

We investigated magnetization dynamics at elevated temperature in terms of a spin model where the model parameters are derived from first principles. Choosing bcc Fe as an example, we focus on different kinds of model assumptions, with temperature-dependent spin model parameters. 
Under equilibrium conditions, the Curie temperature is clearly affected by the different model assumptions, since the values of the exchange constants vary with both the electron temperature and the value of the order parameter assumed in the first-principles calculations. Consequently, the dynamics response of the magnetization to an ultra-short laser pulse can be  affected as well, when the laser power is sufficiently large to reach high electron temperatures and large degrees of demagnetization.

\ack 
This work has been funded by the European Community's Seventh Framework Programme FP7/2007-2013 under grant agreements No.\ 281043, FEMTOSPIN and by the National Research, Development and Innovation Office of Hungary under project No.\ K115575.


\section*{References}

\begin{thebibliography}{10}

\bibitem{LiechtensteinJMMM87}
A.I. Liechtenstein, M.I. Katsnelson, V.P. Antropov, and V.A. Gubanov.
\newblock {\em J. Magn. Magn. Mat.}, 67:65, 1987.

\bibitem{KatsnelsonPRB2000}
M.~I. Katsnelson and A.~I. Lichtenstein.
\newblock {\em Phys. Rev. B}, 61:8906--8912, Apr 2000.

\bibitem{KatsnelsonEPJB2002}
M.~I. Katsnelson and A.~I. Lichtenstein.
\newblock {\em Eur. Phys. J. B}, 30(1):9--15, 2002.

\bibitem{udvardiPRB03}
L.~Udvardi, L.~Szunyogh, K.~Palot{\'a}s, and P.~Weinberger.
\newblock {\em Phys. Rev. B}, 68:104436, 2003.

\bibitem{ebertPRB09}
H.~Ebert and S.~Mankovsky.
\newblock {\em Phys. Rev. B}, 79:045209, Jan 2009.

\bibitem{KatsnelsonPRB2010}
M.~I. Katsnelson, Y.~O. Kvashnin, V.~V. Mazurenko, and A.~I. Lichtenstein.
\newblock {\em Phys. Rev. B}, 82:100403, Sep 2010.

\bibitem{SecchiAnnPhys2015}
A.~Secchi, A.I. Lichtenstein, and M.I. Katsnelson.
\newblock {\em Annals of Physics}, 360:61 -- 97, 2015.

\bibitem{SecchiAnnPhys2013}
A.~Secchi, S.~Brener, A.I. Lichtenstein, and M.I. Katsnelson.
\newblock {\em Annals of Physics}, 333:221 -- 271, 2013.

\bibitem{nowakBOOK07}
U.~Nowak.
\newblock {\em Classical Spin Models}.
\newblock John Wiley \& Sons Ltd., Chichester, 2007.

\bibitem{chimataPRL12}
R.~Chimata, A.~Bergman, L.~Bergqvist, B.~Sanyal, and O.~Eriksson.
\newblock {\em Phys. Rev. Lett.}, 109:157201, 2012.

\bibitem{szilvaPRL13}
A.~Szilva, M.~Costa, A.~Bergman, L.~Szunyogh, L.~Nordstr{\"o}m, and
  O.~Eriksson.
\newblock {\em Phys. Rev. Lett.}, 111:127204, Sep 2013.

\bibitem{bottcherJMMM12}
D~B{\"o}ttcher, A~Ernst, and J~Henk.
\newblock {\em J. Magn. Magn. Mat.}, 324:610, 2012.

\bibitem{gyorffyJPFMF85}
B.~L. Gy{\"o}rffy, A.~J. Pindor, J.~B. Staunton, G.~M. Stocks, and H.~Winter.
\newblock {\em J. Phys. F: Met. Phys.}, 15:1337, 1985.

\bibitem{stauntonPRL04}
J.~B. Staunton, S.~Ostanin, S.~S.~A. Razee, B.~L. Gy{\"o}rffy, L.~Szunyogh,
  B.~Ginatempo, and E.~Bruno.
\newblock {\em Phys. Rev. Lett.}, 93:257204, 2004.

\bibitem{stauntonPRB06}
J.~B. Staunton, L.~Szunyogh, A.~Buruzs, B.~L. Gy{\"o}rffy, S.~Ostanin, and
  L.~Udvardi.
\newblock {\em Phys. Rev. B}, 74:144411, 2006.

\bibitem{beaurepairePRL96}
E.~Beaurepaire, J.-C. Merle, A.~Daunois, and J.~Y. Bigot.
\newblock {\em Phys. Rev. Lett.}, 76:4250, 1996.

\bibitem{KKRBOOK05}
J.~Zabloudil, R.~Hammerling, L.~Szunyogh, and P.~Weinberger.
\newblock {\em Electron Scattering in Solid Matter}.
\newblock Springer, Heidelberg, 2005.

\bibitem{deakPRB14}
A.~De{\'a}k, E.~Simon, L.~Balogh, L.~Szunyogh, M.~dos Santos~Dias, and J.~B.
  Staunton.
\newblock {\em Phys. Rev. B}, 89:224401, 2014.

\bibitem{Mermin-1965}
N.~David Mermin.
\newblock {\em Phys. Rev.}, 137:A1441--A1443, Mar 1965.

\bibitem{KueblerBook}
J.~K{\"u}bler.
\newblock {\em Theory of Itinerant Electron Magnetism}.
\newblock Oxford University Press, Oxford, 2009.

\bibitem{Gunnarsson-1976}
O~Gunnarsson.
\newblock {\em Journal of Physics F: Metal Physics}, 6(4):587, 1976.

\bibitem{szunyoghPRB11}
L.~Szunyogh, L.~Udvardi, J.~Jackson, U.~Nowak, and R.~Chantrell.
\newblock {\em Phys. Rev. B}, 83:024401, 2011.

\bibitem{Butler-1985}
W.~H. Butler.
\newblock {\em Phys. Rev. B}, 31:3260--3277, Mar 1985.

\bibitem{hinzkeJMMM00}
D.~Hinzke and U.~Nowak.
\newblock {\em J. Magn. Magn. Mat.}, 221:365, 2000.

\bibitem{mjw-book-1978}
V.~L. Moruzzi, J.~F. Janak, and A.~R. Williams.
\newblock {\em Calculated Electronic Properties of Metals}.
\newblock Pergamon, New York, 1978.

\bibitem{Hunter-2007}
J.~D. Hunter.
\newblock {\em Computing In Science \& Engineering}, 9(3):90--95, 2007.

\bibitem{Buruzs-2008}
{\'A}.~Buruzs, L.~Szunyogh, and P.~Weinberger.
\newblock {\em Philos. Mag.}, 88(18-20):2615--2626, 2008.

\bibitem{huhnePRB98}
T.~Huhne, C.~Zecha, H.~Ebert, P.~H. Dederichs, and R.~Zeller.
\newblock {\em Phys. Rev. B}, 58:10236--10247, Oct 1998.

\bibitem{kazantsevaEPL08}
N.~Kazantseva, U.~Nowak, R.~W. Chantrell, J.~Hohlfeld, and A.~Rebei.
\newblock {\em Europhys. Lett.}, 81:27004, 2008.

\bibitem{hinzkePRB15}
D.~Hinzke, U.~Atxitia, K.~Carva, P.~Nieves, O.~Chubykalo-Fesenko, P.~M.
  Oppeneer, and U.~Nowak.
\newblock {\em Phys. Rev. B}, 92:054412, 2015.

\bibitem{kaganovJETP57}
M.~I. Kaganov, I.~M. Lifshitz, and L.~V. Tanatarov.
\newblock {\em Sov. Phys. JETP}, 4:173, 1957.

\end{thebibliography}
\end{document}